\documentclass{article}

\usepackage{arxiv}

\usepackage[utf8]{inputenc} 
\usepackage[T1]{fontenc}    
\usepackage{hyperref}       
\usepackage{url}            
\usepackage{booktabs}       
\usepackage{amsfonts}       
\usepackage{nicefrac}       
\usepackage{microtype}      
\usepackage{lipsum}		
\usepackage{graphicx}
\usepackage{natbib}
\usepackage{doi}

\usepackage{amsmath} 

\title{Towards Sustainable Magnetic Resonance Imaging: Insights from long-term, high-resolution energy recordings across an entire scanner fleet}

\author{
\href{https://orcid.org/0009-0000-7551-2770}{\includegraphics[width=0.7em]{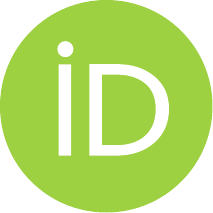}\hspace{1mm}Florian Leonhard Raab}$^{1,2,3}$,
\href{https://orcid.org/0000-0002-7441-1406}{\includegraphics[width=0.7em]{orcid.pdf}\hspace{1mm}Fiona Mankertz}$^2$,
\href{https://orcid.org/0000-0003-4547-4229}{\includegraphics[width=0.7em]{orcid.pdf}\hspace{1mm}Nour Maalouf}$^2$,
\href{https://orcid.org/0009-0004-7149-0323}{\includegraphics[width=0.7em]{orcid.pdf}\hspace{1mm}Josephine Berger}$^2$,
\href{https://orcid.org/0000-0001-8583-8627}{\includegraphics[width=0.7em]{orcid.pdf}\hspace{1mm}Andreas Lingg}$^2$,
\And
\href{https://orcid.org/0000-0002-3496-8808}{\includegraphics[width=0.7em]{orcid.pdf}\hspace{1mm}Reza Dehdab}$^2$,
\href{https://orcid.org/0000-0003-4542-6462}{\includegraphics[width=0.7em]{orcid.pdf}\hspace{1mm}Sebastian Werner}$^2$,
\href{https://orcid.org/0000-0003-1535-8980}{\includegraphics[width=0.7em]{orcid.pdf}\hspace{1mm}Judith Herrmann}$^2$,
\href{https://orcid.org/0000-0002-5961-9727}{\includegraphics[width=0.7em]{orcid.pdf}\hspace{1mm}Andreas Brendlin}$^2$,
\href{https://orcid.org/0000-0003-3701-2227}{\includegraphics[width=0.7em]{orcid.pdf}\hspace{1mm}Sebastian Gassenmaier}$^2$,
\And
Suhas Siddaramu$^{1,2}$,
\href{https://orcid.org/0000-0002-5964-6485}{\includegraphics[width=0.7em]{orcid.pdf}\hspace{1mm}Fabian Wagner}$^4$,
Julian Wohlers$^4$,
Shreeja Varadarajan$^4$,
Gurlal Singh$^4$,
\And
Jens Gühring$^4$,
Rainer Schneider$^4$,
Konstantin Nikolaou$^2$,
\href{https://orcid.org/0000-0002-6861-6245}{\includegraphics[width=0.7em]{orcid.pdf}\hspace{1mm}Saif Afat}$^2$,
\href{https://orcid.org/0000-0002-0353-4898}{\includegraphics[width=0.7em]{orcid.pdf}\hspace{1mm}Thomas Küstner}$^{1,2,3}$\\[12pt]
\small
$^1$Medical Image and Data Analysis (MIDAS.lab), Tübingen University Hospital\\
$^2$Department of Diagnostic and Interventional Radiology, Tübingen University Hospital\\
$^3$Faculty of Computer Science, Eberhard Karls University of Tübingen\\
$^4$Magnetic Resonance, Siemens Healthineers AG
}

\renewcommand{\shorttitle}{Towards Sustainable MRI}

\hypersetup{
pdftitle={Towards Sustainable Magnetic Resonance Imaging: Insights from long-term, high-resolution energy recordings across an entire scanner fleet},
pdfsubject={cs.OH},
pdfauthor={Florian Leonhard Raab},
pdfkeywords={MRI energy consumption, Healthcare sustainability, Data-driven analysis, MRI data integration for energy analysis, Live MRI energy monitoring},
}
\begin{document}
\maketitle

\begin{abstract}
Magnetic resonance imaging (MRI) is among the most energy-intensive diagnostic modalities in healthcare, yet its energy consumption and the factors influencing it remain insufficiently understood. This study aims to establish a comprehensive baseline of MRI energy consumption by characterizing energy demand across a scanner fleet, examining scanner utilization and operating patterns in clinical practice. Concurrently, it investigates the relationships between energy consumption and relevant operational and acquisition features. \\

On average, a single MRI measurement consumed 0.43 kWh, while a complete examination consumed 13.50 kWh. In general, substantial differences in energy consumption were observed between  MRI scanners and their corresponding operating modes (scan, idle, and eco-power mode). These variations may be related to differences in scanner operating patterns, employed examination protocols, and their resulting acquisition parameters. Idle and eco-power modes accounted for more cumulative energy consumption than active scanning. However, these energy shares should always be interpreted in relation to scanner occupancy, as utilization patterns strongly influence the distribution of energy across those operating modes. Lastly, linear regression analysis showed that energy consumption was more strongly associated with scan duration than with average power, suggesting that scan duration may be an important factor influencing MRI energy consumption. \\

To support the analysis of the complex dataset and integrate the different analytical steps, an interactive dashboard was developed to enable systematic data exploration and facilitate the identification of energy consumption patterns and relationships across different analysis stages. In addition to preprocessed data, the dashboard incorporates live data streamed directly from the hospital, enabling continuous monitoring of scanner power consumption. Overall, this study provides a comprehensive characterization of MRI energy consumption across multiple levels of analysis, establishing a foundation for future data-driven modeling and optimization approaches. \\
\end{abstract}


\section{Introduction}
\label{introduction}
The healthcare sector is a major consumer of energy and a significant contributor to global $CO_2$ emissions, corresponding to about 4.4\% of global net emissions \cite{Bosurgi_2019, Keil_2024, Lenzen_2020, Karliner_2019}. Despite this considerable environmental impact, sustainability has historically received limited attention within healthcare, where the primary focus has naturally been on improving patient care and advancing medical technologies \cite{Sherman_2020, Reising_2025}. As a result, sustainability was not integrated as a core objective when many healthcare facilities and systems were designed, rendering its implementation more challenging today. \\

At the same time, the energy-intensive nature of hospitals leaves them particularly vulnerable to fluctuations in energy markets \cite{Bawaneh_2019}. In recent years, uncertainties in European energy production and energy trade have led to an increase in energy prices \cite{EU_2024, EU_2025}, placing additional financial pressure on healthcare providers. Consequently, hospitals are faced with rising operating costs while also being expected to reduce their environmental footprint. This combination of economic and environmental pressures highlights the growing need to integrate sustainability into hospital operations. \\

Many conventional approaches to improving sustainability, such as heating, ventilation, and air conditioning (HVAC) systems, are associated with high financial costs, lengthy implementation times, and considerable practical constraints, especially in older hospital infrastructures. Yet, radiology departments offer a promising opportunity to optimize energy efficiency, as they are among the most energy-intensive areas within a hospital \cite{Villa_2021}. This high energy demand is primarily driven by the operation of imaging systems such as Magnetic Resonance Imaging (MRI), Computed Tomography (CT), and Ultrasound (US) devices \cite{Heye_2020}. Collectively, these systems can account for up to 7.3\% of a hospital's total electricity consumption, with MRI scanners representing the largest share of energy use among imaging modalities \cite{Villa_2021}. Consequently, improving the energy efficiency of MRI devices presents a significant opportunity to reduce the hospital’s overall energy demand. Moreover, MRI scanners can be optimized through sequence design and software-based solutions, avoiding the need for major structural modifications or substantial financial investments \cite{Doo_2024}. \\

Despite growing interest in healthcare sustainability, existing studies on MRI energy consumption have largely examined individual aspects of scanner operation or specific energy-saving interventions, rather than systematically linking energy demand across different levels of operations. \citet{Heye_2020} characterized MRI energy consumption according to body region and scanner operating mode, distinguishing between scanning, idle, and eco-power modes, and quantified potential energy and cost savings associated with changes in scanner operation. Similarly, \citet{Heye_2023} focused on the energy consumed by nonproductive imaging equipment during periods of inactivity and demonstrated the potential for reducing energy consumption by switching off devices when they were not required. Extending this operational perspective, \citet{Woolen_2023} investigated different scanner power modes and operational strategies to reduce energy consumption and the associated carbon footprint. \citet{Afat_2025} examined accelerated musculoskeletal MRI protocols using Deep Learning (DL) methods and optimized magnet-cooling patterns to reduce energy consumption. Collectively, these studies demonstrate multiple opportunities for reducing MRI energy consumption, but primarily consider individual operational factors or interventions in isolation. \\

A comprehensive characterization that connects scanner-level energy demand and utilization with examination characteristics, scanning protocols, and acquisition parameters is therefore warranted. To address this gap, we analyze energy consumption and scanner utilization across a fleet of seven MRI scanners using data sampled at one-second intervals over 1.5 years. By integrating energy recordings with operational, examination, and acquisition data, this approach establishes a robust baseline of MRI energy consumption in a hospital setting while providing a more complete understanding of the factors driving energy demand. These findings provide a foundation for future research on effective energy-saving strategies. \\

\section{Methods}
\label{methods}
This retrospective observational study analyzed the energy consumption and usage of seven MRI scanners at the University Hospital Tübingen, Germany. The study was approved by the Ethics Committee of the Faculty of Medicine, Eberhard Karls University Tübingen, and University Hospital Tübingen (Institutional Review Board No. 015/2026A)

\subsection{Data Collection and Preprocessing}
\label{data_collection_preprocessing}

\begin{figure}[t]
    \centering
    \includegraphics[
        width=0.8 \linewidth,
        keepaspectratio
    ]{figure1_data_sampling_merging.pdf}
    \caption{An overview of the data processing architecture. Power meters recorded the energy and power consumption of the MRI scanner and its corresponding reconstruction workstation, while MRI system log files were acquired. These complementary data sources were subsequently integrated into a unified dataset.}
    \label{fig:figure1_data_sampling_merging}
\end{figure}

 To generate a coherent dataset, electrical energy consumption and MRI system log data were collected and subsequently integrated (Fig \ref{fig:figure1_data_sampling_merging}). Data were collected from four 1.5-T MRI systems (2x MAGNETOM Aera, MAGNETOM AvantoFit, and MAGNETOM Sola) and three 3-T systems (MAGNETOM PrismaFit, MAGNETOM Vida, and MAGNETOM VidaFit). For brevity and clarity, scanner models are referred to throughout this study by the final component of the model designation, following the manufacturer’s nomenclature (Siemens Healthineers, Erlangen, Germany). The energy consumption of each MRI scanner (complete system including its corresponding reconstruction workstation) was measured using Siemens PAC3200T power meters with a sampling rate of 1 Hz, which are installed at each scanner site. In parallel, MRI system log files were acquired, which contained information on scanner operational states, scanning activity, system events, and scan parameters. \\

The energy recordings and MRI system logs were temporally synchronized and merged based on their respective timestamps. The total energy consumption was calculated for each examination and each individual measurement. Each examination consists of multiple measurements, referred to as sequences, which include localizers, automatic adjustment scans, and diagnostic imaging sequences. In addition, the scanner's total daily energy consumption was determined and categorized into energy consumed during eco-power mode, idle mode, and scan mode. Eco-power mode refers to the vendor-defined energy-saving mode used when the scanner is not in use. In this mode, the helium pump, which is required to cool the system, is switched off and automatically reactivated when predefined temperature or pressure thresholds are reached. Scan mode was defined as the period during which the scanner was actively acquiring images. The remaining idle and eco-power modes were differentiated based on predefined power thresholds. Periods with power consumption above the threshold were classified as idle mode, while those below the threshold were classified as eco-power mode. \\

The resulting dataset comprised metadata describing each employed MRI scanner, information on each conducted examination, and detailed records for each measurement performed within an examination, including the corresponding acquisition parameters and the calculated energy consumption. In total, 26,471 files encompassing 68 GB were combined to create a coherent dataset, which spanned from 01.01.2024 to 30.11.2025. \\

Before the analysis, the dataset underwent several filtering steps to remove implausible observations. Initially, the dataset comprised 600,293 individual measurements. Entire examinations with a duration exceeding 3 hours were excluded, resulting in the removal of 359 measurements. Subsequently, individual measurements with a duration greater than 30 minutes were discarded, resulting in the removal of an additional 402 measurements. This was performed to ensure the filtering out of unrealistically long examinations or measurements. These thresholds were established based on domain knowledge provided by the MTAs.

\subsection{Data Analysis}
\label{data_analysis}
An overview of the daily number of examinations and measurements was visualized for all MRI scanners using stacked bar plots. In addition, the corresponding energy consumption associated with each examination and measurement was displayed. \\

Scanner utilization was evaluated by calculating the occupancy rate of each MRI scanner. Occupancy was defined as the proportion of the sum of total daily scanning time relative to the total duration of a day ($24 \text{ hours} \cdot 60 \text{ minutes} \cdot 60 \text{ seconds}$), with both quantities expressed in seconds. The occupancy rate was calculated for each day and subsequently averaged across the study period to obtain the mean occupancy for each scanner. \\

To characterize temporal variations in energy demand, the energy consumption of each scanner was averaged by weekday and evaluated across the three operating modes (eco-power, idle, and scan). Additionally, overall fleet energy demand was assessed by aggregating energy consumption across all scanners for each operating mode. For both analyses, energy consumption was first averaged separately for each scanner, as data availability varied between machines. These scanner-specific values were subsequently aggregated to obtain more robust estimates of energy consumption across the fleet. \\

Based on the fundamental relationship between energy, power, and time ($E = P \cdot t$), we hypothesized that scan duration and the average power consumption are the primary determinants of energy consumption. To investigate the influence of these factors, measurements were grouped by sequence type, and the average energy consumption was calculated for each unique sequence. Scatter plots with linear regression models were then used to evaluate the relationships between sequence-specific energy consumption and scan duration, as well as between sequence-specific energy consumption and average power consumption. \\

Last, due to the complexity of the dataset, an interactive dashboard was developed to facilitate data exploration. The dashboard and associated visualizations enable users to filter and examine selected data points and specific regions of interest. Interactive hover functionality provides additional information for individual data points. Furthermore, the dashboard incorporates a live data connection to enable continuous monitoring and includes additional visualizations to support a comprehensive analysis. \\

For the visualization of energy consumption, error bars were not included, as they were not compatible with the intended visualization of the stacked bar plots and pie charts. This was an intentional trade-off to allow a more granular differentiation of energy consumption at the individual scanner level. Furthermore, the energy and power values were derived from log data rather than repeated recordings subject to measurement noise. Therefore, measures of statistical variability would not directly represent the accuracy or uncertainty of the underlying recordings and could be misleading. \\

Where a distribution of observations was available, descriptive statistics including the mean, median, first quartile (Q1), and third quartile (Q3) were calculated to provide an overall characterization of the data and its variability. These statistics were only calculated when the underlying distribution was retained. In analyses where observations were first averaged at the scanner level and subsequently aggregated across scanners, the resulting values represented scanner-level averages rather than a distribution of individual observations. In such cases, median and quartile estimates were not calculated, as they would not provide a meaningful representation of the underlying data distribution. \\

\section{Results}
\label{results}

\begin{figure}[t]
    \centering
    \includegraphics[
        width=\linewidth,
        keepaspectratio
    ]{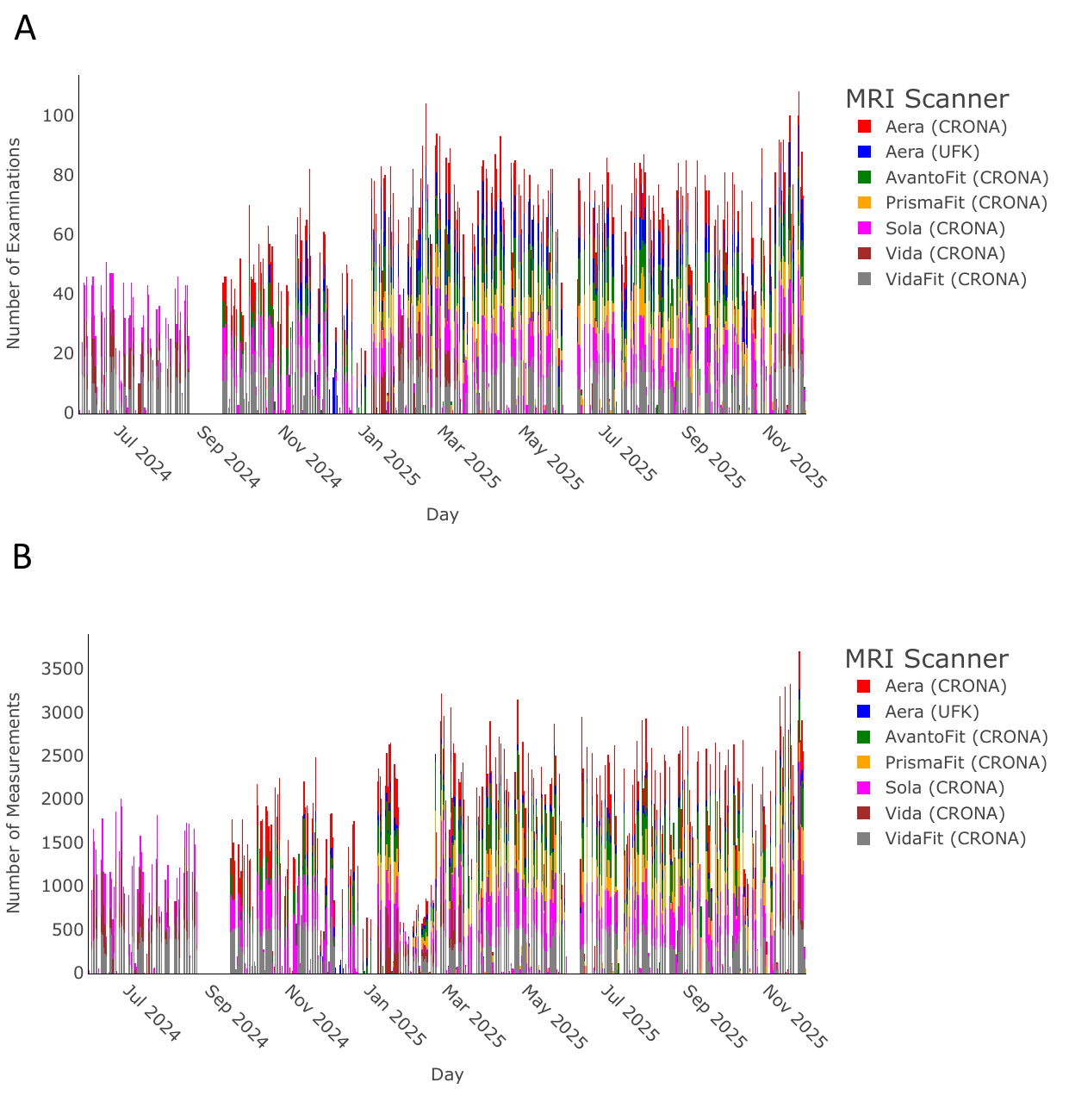}
    \caption{Daily scanner-specific examination and measurement volumes over the study period. MRI scanners are color-coded for comparison. Seven MRI scanners across two sites (CRONA and UFK) were consecutively onboarded. On average, 42.80 examinations and 1,329.34 measurements were performed per day. Inter-scanner differences in workload and periods of missing data due to data recording interruptions can be seen. (A) Average daily number of MRI examinations. (B) Average daily number of MRI measurements.}
    \label{fig:figure2_examinations_measurement_count}
\end{figure}

Seven MRI scanners across two sites, designated CRONA and UFK, were consecutively onboarded. The Sola (CRONA), Vida (CRONA), and VidaFit (CRONA) scanners were onboarded on 01.06.2024, followed by the AvantoFit (CRONA) and Aera (CRONA) scanners on 17.09.2024. Subsequently, the Aera (UFK) scanner was included on 22.10.2024, and the PrismaFit (CRONA) scanner was added on 07.01.2025. On average, 42.8 examinations (Median: 44 [Q1: 13, Q3: 56]) and 1329.3 measurements (Median: 1351 [Q1: 337, Q3: 1802]) were performed per day (Fig. \ref{fig:figure2_examinations_measurement_count}). Notable inter-scanner differences in both the number of examinations and the number of measurements were observed. The high measurement count is primarily driven by the inclusion of adjustment and localizer scans, which are counted as individual measurements. Additionally, periods with missing data due to data transfer interruptions became evident. \\

\begin{figure}[t]
    \centering
    \includegraphics[
        width=\linewidth,
        keepaspectratio
    ]{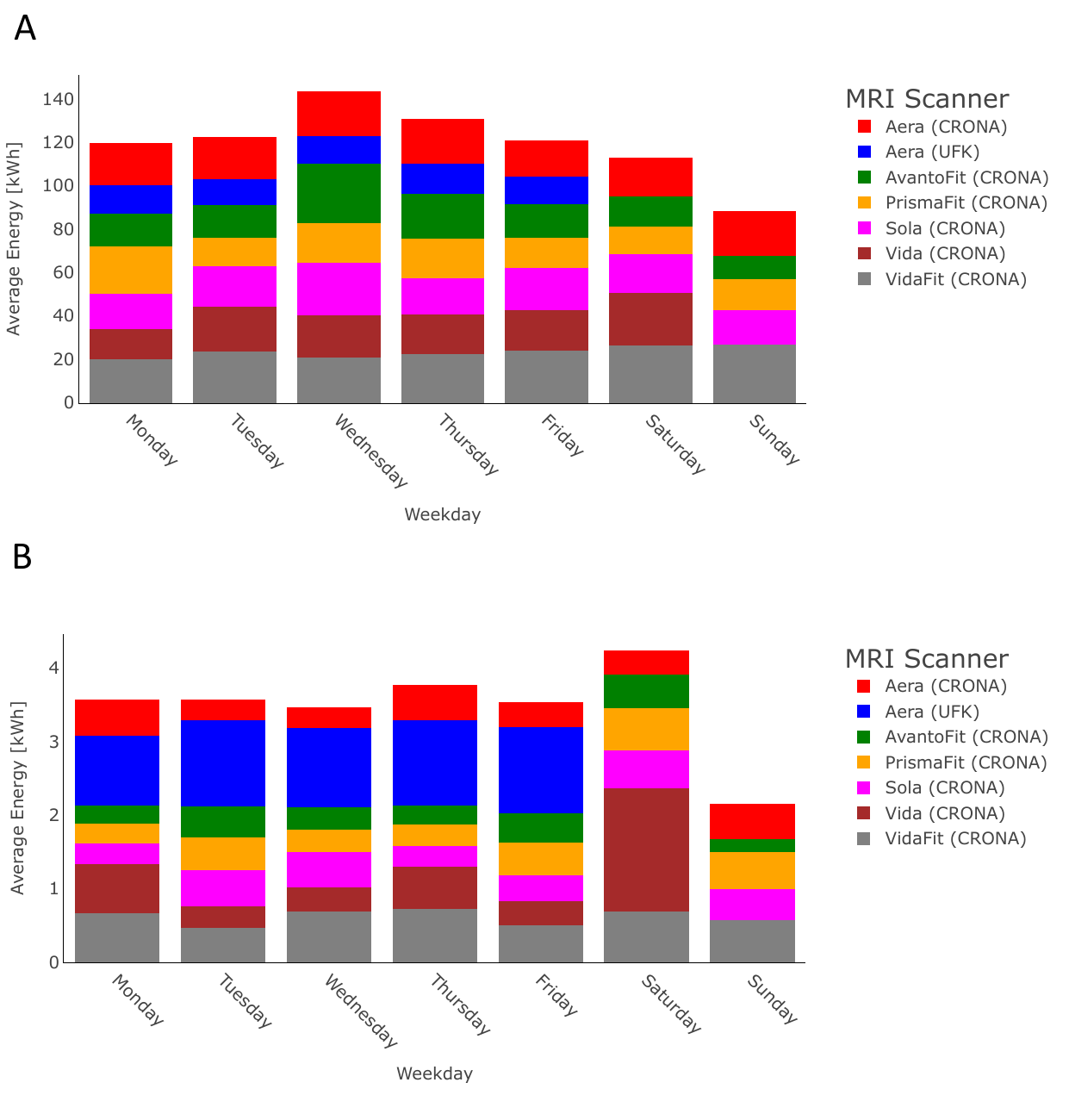}
    \caption{Scanner-specific MRI energy consumption aggregated across weekdays. Average energy consumption amounted to 13.02 kWh per examination and 0.50 kWh per measurement. Energy consumption varied considerably between MRI scanners, reflecting differences in scanner characteristics, applied protocols, and the number of measurements. Temporal differences reflected scanner-specific operating schedules, with reduced consumption during weekends. (A) Average energy consumption per MRI examination. (B) Average energy consumption per MRI measurement.}
    \label{fig:figure3_energy_protocols_sequences}
\end{figure}

On average, a single MRI measurement consumed 0.43 kWh (Median: 0.04 kWh [Q1: 0.01  kWh, Q3: 0.25  kWh]), while a complete examination consumed 13.50  kWh (Median: 11.96 kWh [Q1: 8.42  kWh, Q3: 16.94  kWh]) per scanner. Temporal fluctuations in total energy consumption were relatively modest. However, energy consumption decreased markedly on Sundays. Whereas the Vida (CRONA) was operated only on Saturdays but not on Sundays, contrary to the Aera (UFK), which was not operated during weekends (Fig. \ref{fig:figure3_energy_protocols_sequences}). \\

Distinct inter-scanner differences in energy consumption were evident, with three groups emerging based on examination-level energy use. The Aera (UFK), and PrismaFit (CRONA) exhibited the lowest average energy consumption per examination, at 9.70 kWh, and 11.44 kWh, respectively. Intermediate values were observed for the Aera (CRONA), Sola (CRONA), Vida (CRONA), and AvantoFit (CRONA), each consuming approximately 13.7 kWh per examination. In contrast, the VidaFit (CRONA) showed the highest average energy consumption, at 17.51 kWh per examination (Fig. \ref{fig:figure3_energy_protocols_sequences} A). \\

A slightly different pattern emerged when energy consumption was analyzed at the measurement level. The Aera (CRONA), Sola (CRONA), PrismaFit (CRONA), Vida (CRONA), and AvantoFit (CRONA) each consumed approximately 0.36 kWh per measurement on average. By comparison, the VidaFit (CRONA) and Aera (UFK) exhibited substantially higher values, averaging 0.60 kWh and 1.10 kWh per measurement (Fig. \ref{fig:figure3_energy_protocols_sequences} B). \\

\begin{figure}[t]
    \centering
    \includegraphics[
        width=\linewidth,
        keepaspectratio
    ]{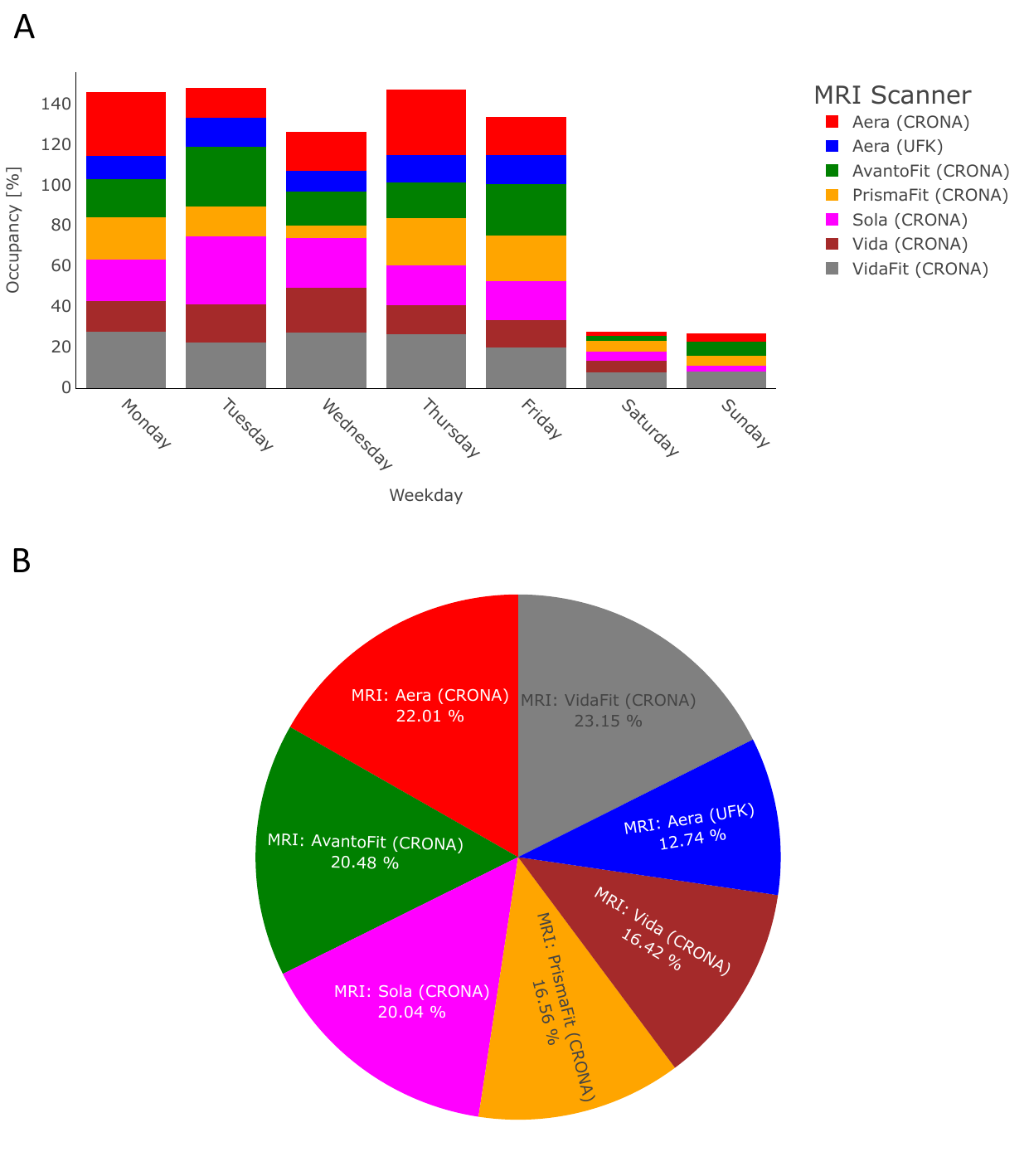}
    \caption{Scanner-specific occupancy. Scanner occupancy was defined as the cumulative daily scan time relative to the total time per day ($24 \text{ hours} \cdot 60 \text{ minutes} \cdot 60 \text{ seconds}$). (A) Mean occupancy by weekday demonstrates temporal variations in scanner utilization and reduced occupancy during weekends. (B) Mean occupancy by scanner over the study period. Averaged Pie chart slice sizes are proportional to the average occupancy of each scanner. Differences in scanner occupancy indicate variations in workload distribution and patient throughput, which influence energy consumption patterns.}
    \label{fig:figure4_occupancy_scanner}
\end{figure}

Across all scanners, the mean scanner occupancy was 19.38\% (Median: 20.04 \% [Q1: 16.56\%, Q3: 22.01\%]). Occupancy decreased markedly during weekends, whereas only minor day-to-day variation was observed during weekdays. Especially, the PrismaFit (CRONA) showed the largest weekday variation, with occupancy ranging from 11.87\% to 32.42\%. In contrast, occupancy for the remaining scanners remained relatively stable throughout the week. The occupancy analysis further highlighted scanner-specific differences in weekend operation: the Vida (CRONA) was operated on Saturdays but not on Sundays, whereas the Aera (UFK) was not utilized during weekends (Fig. \ref{fig:figure4_occupancy_scanner} A). \\

When occupancy was analyzed by scanner, three distinct utilization groups emerged. The Aera (UFK) exhibited the lowest average occupancy (12.74\%). The Vida (CRONA) and PrismaFit (CRONA) formed an intermediate group, with an average occupancy of approximately 16.49\%. In contrast, the VidaFit (CRONA), Aera (CRONA), AvantoFit (CRONA), and Sola (CRONA) exhibited the highest occupancy, ranging from 20.04\% to 23.15\% (Fig. \ref{fig:figure4_occupancy_scanner} B). \\

\begin{figure}[t]
    \centering
    \includegraphics[
        width=\linewidth,
        keepaspectratio
    ]{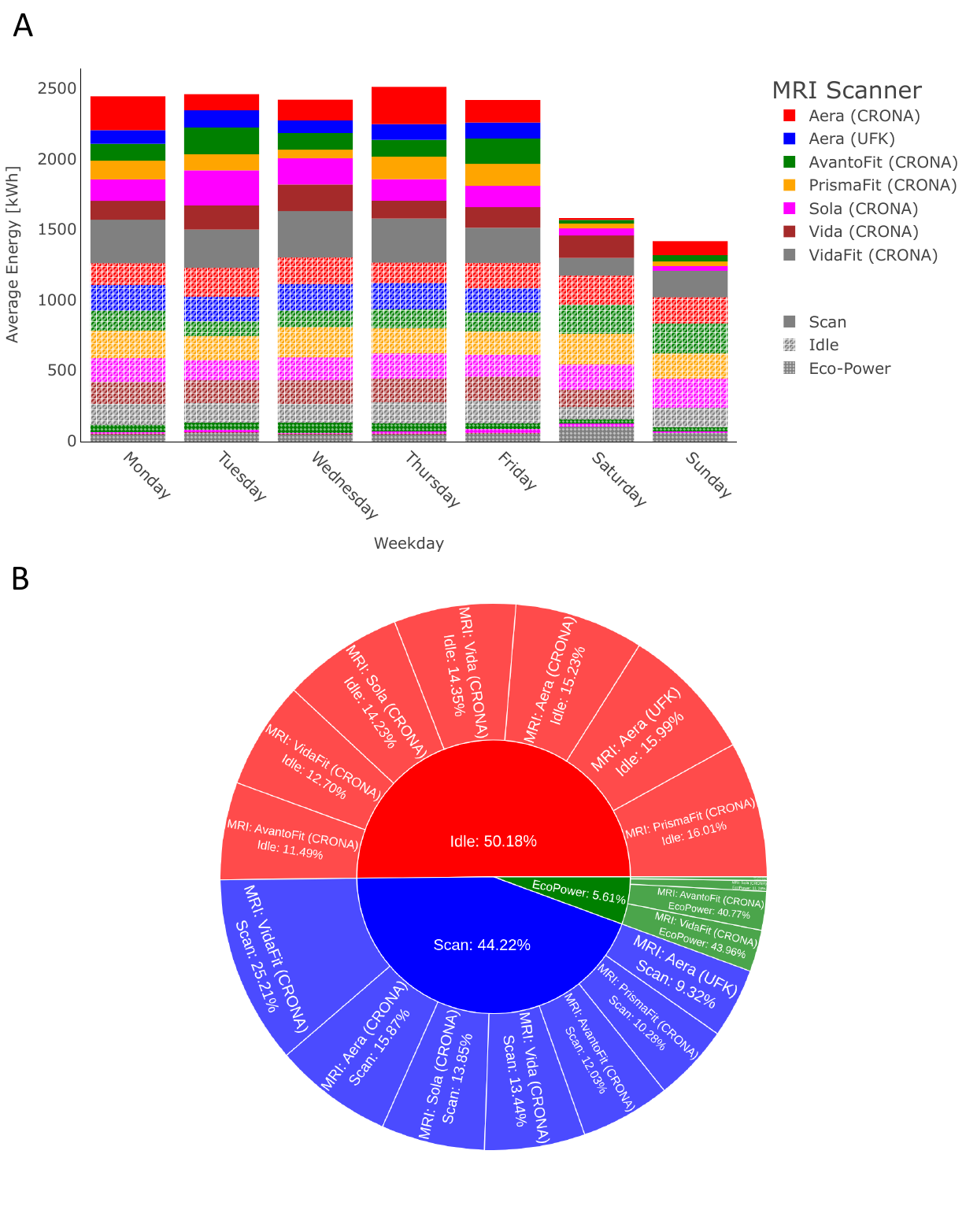}
    \caption{(A) Average scanner-specific energy consumption by operating mode and weekday. Total energy consumption is shown for each MRI scanner, and fill patterns indicate the operating mode: solid bars represent scan mode, striped bars represent idle mode, and dotted bars represent eco-power mode. Temporal variations in energy consumption across weekdays and weekends were observed, primarily driven by differences in scan-mode utilization, while idle and eco-power energy consumption remained relatively stable due to continuous cooling requirements. (B) Average scanner-specific energy consumption by operating mode. Energy consumption is expressed as the proportion of the total energy consumption for each scanner. Blue slices represent scan mode, red slices represent idle mode, and green slices represent eco-power mode. Idle and eco-power modes cumulatively accounted for a larger proportion of total energy consumption than active scanning, highlighting the substantial contribution of non-scanning periods to overall MRI energy demand.}
    \label{fig:figure5_scan_idle_ecopower_energy}
\end{figure}

Energy consumption during idle and eco-power modes remained relatively stable across weekdays and weekends, whereas energy consumption during active scanning varied according to scanner utilization and day of the week. During weekdays, total daily energy consumption ranged from 2420.95 to 2511.21 kWh, comprising scan-mode consumption ranging from 1118.86 to 1245.49 kWh, idle-mode consumption ranging from 1094.35 to 1168.87 kWh, and eco-power consumption ranging from 119.19 to 134.92 kWh. Weekend energy consumption was substantially lower, with a total consumption of 1581.34 kWh on Saturdays and 1417.86 kWh on Sundays. On Saturdays, the total energy consumption comprised 406.78 kWh during scanning, 1015.50 kWh during idling, and 159.07 kWh in eco-power mode. On Sundays, total energy consumption comprised 396.50 kWh during scan mode, 923.02 kWh during idle mode, and 98.33 kWh during eco-power mode (Fig. \ref{fig:figure5_scan_idle_ecopower_energy} A). \\

Temporal aggregation of energy consumption across all scanners and their respective operating modes revealed distinct patterns. On average, the entire scanner fleet consumed 2354.89 kWh per day, comprising 1041.27 kWh during scanning, 1181.60 kWh while idling, and 132.00 kWh in eco-power mode. On a per-scanner basis, the average daily energy consumption amounted to 343.42 kWh (Median: 366.86 kWh [Q1: 289.41 kWH, Q3: 390.17 kWh]), consisting of 154.08 kWh during scanning (Median: 175.29 kWh [Q1: 115.04 kWH, Q3: 208.41 kWh]), 164.92 kWh while idling (Median: 164.92 kWh [Q1: 148.58 kWH, Q3: 186.97 kWh]), and 24.42 kWh in eco-power mode (Median: 20.41 kWh [Q1: 9.98 kWH, Q3: 30.77 kWh]) (Fig. \ref{fig:figure5_scan_idle_ecopower_energy} B). Importantly, in the analyses described above, energy consumption was first averaged at the scanner level to account for missing or incomplete data. Consequently, these analyses were based on scanner-level averages rather than the underlying distribution of individual measurements, making it inappropriate to calculate meaningful medians and quantiles for the entire scanner fleet. \\

Analysis of energy consumption during active scanning revealed two distinct scanner groups. The VidaFit (CRONA) exhibited substantially higher average scan-mode energy consumption (262.40 kWh) compared with the remaining scanners, which ranged from 97.07 kWh to 165.24 kWh. In contrast, idle-mode energy consumption showed no clear grouping and varied within a narrower range across scanners, from 135.73 kWh to 188.99 kWh. Eco-power mode was utilized by only four scanners: AvantoFit (CRONA), Sola (CRONA), VidaFit (CRONA), and Vida (CRONA). Consumption was highest for the VidaFit (CRONA) and AvantoFit (CRONA) at 58.03 kWh and 53.82 kWh, respectively, while Sola (CRONA) and Vida (CRONA) showed lower consumption at 14.89 kWh and 5.26 kWh, respectively. \\

A general trend was observed in which scanners with higher energy consumption during active scanning tended to exhibit lower energy consumption during idle and eco-power modes. However, this relationship was not consistent across all scanners and did not fully account for the observed inter-scanner differences. \\

\begin{figure}[t]
    \centering
    \includegraphics[
        width=\linewidth,
        keepaspectratio
    ]{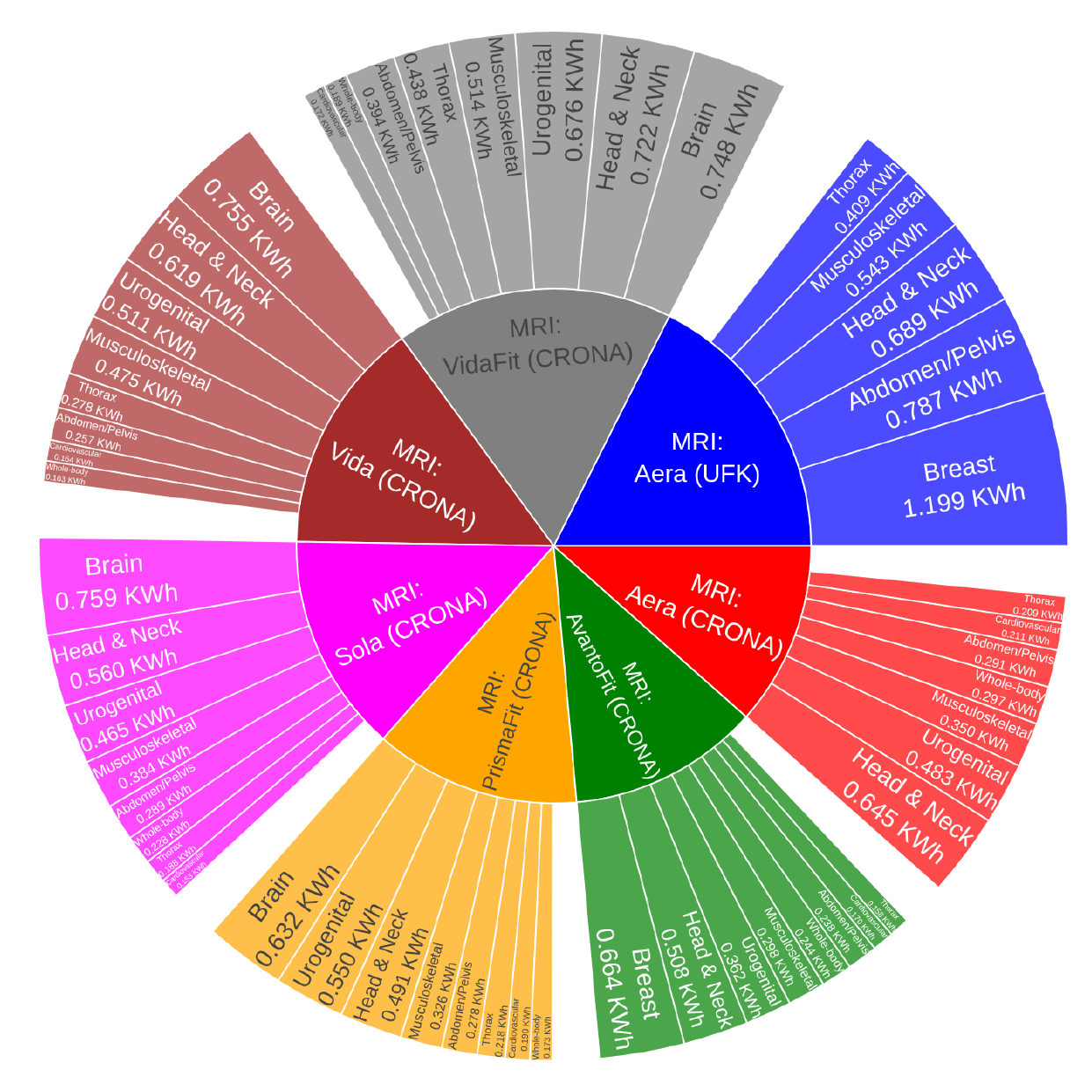}
    \caption{Average scanner-specific energy consumption per measurement grouped by examined body region (i.e. scan mode). Pie charts illustrate the distribution of energy consumption across body regions for each MRI scanner. Energy consumption varied substantially between both body regions and scanners, indicating scanner-specific differences in energy demand across examination types. While body-region-based grouping provides an overview of energy demand across clinical applications, it does not fully account for variations caused by differences in sequences and acquisition parameters, highlighting the need for future sequence-based, data-driven analyses.}
    \label{fig:figure6_energy_scanner_bodyregion_sunburst}
\end{figure}

An overview of energy consumption across clinical applications is illustrated in Fig. \ref{fig:figure6_energy_scanner_bodyregion_sunburst}. The average energy consumption per measurement was grouped according to the examined body region. Distinct differences between body regions were observed. Breast examinations exhibited the highest overall average energy consumption (1.20 kWh per sequence), followed by brain examinations (0.74 kWh), head and neck examinations (0.62 kWh), and urogenital examinations (0.55 kWh). Intermediate energy consumption was observed for musculoskeletal examinations (0.38 kWh) and Abdomino/Pelvis (0.30 kWh). In contrast, whole-body examinations (0.21 kWh), thoracic (0.21 kWh), and cardiovascular (0.18 kWh) exhibited the lowest average energy consumption per sequence. The latter sequences are predominantly respiratory- or cardiac-triggered and/or performed with breath-holding to reduce motion artifacts and improve image quality, which may contribute to shorter acquisition times. \\

In addition to body region-specific differences, scanner-dependent variations in energy consumption were evident. Most notably, breast examinations performed on the Aera (UFK) exhibited substantially higher energy consumption, averaging 1.20 kWh per examination, compared with breast examinations acquired on other scanners, such as the AvantoFit (CRONA), which required only 0.66 kWh per breast measurement. This difference may be partly attributable to the longer acquisition times associated with the dedicated breast protocols used on the Aera (UFK) compared with the screening protocols used on the AvantoFit (CRONA). \\

\begin{figure}[t]
    \centering
    \includegraphics[
        width=\linewidth,
        keepaspectratio
    ]{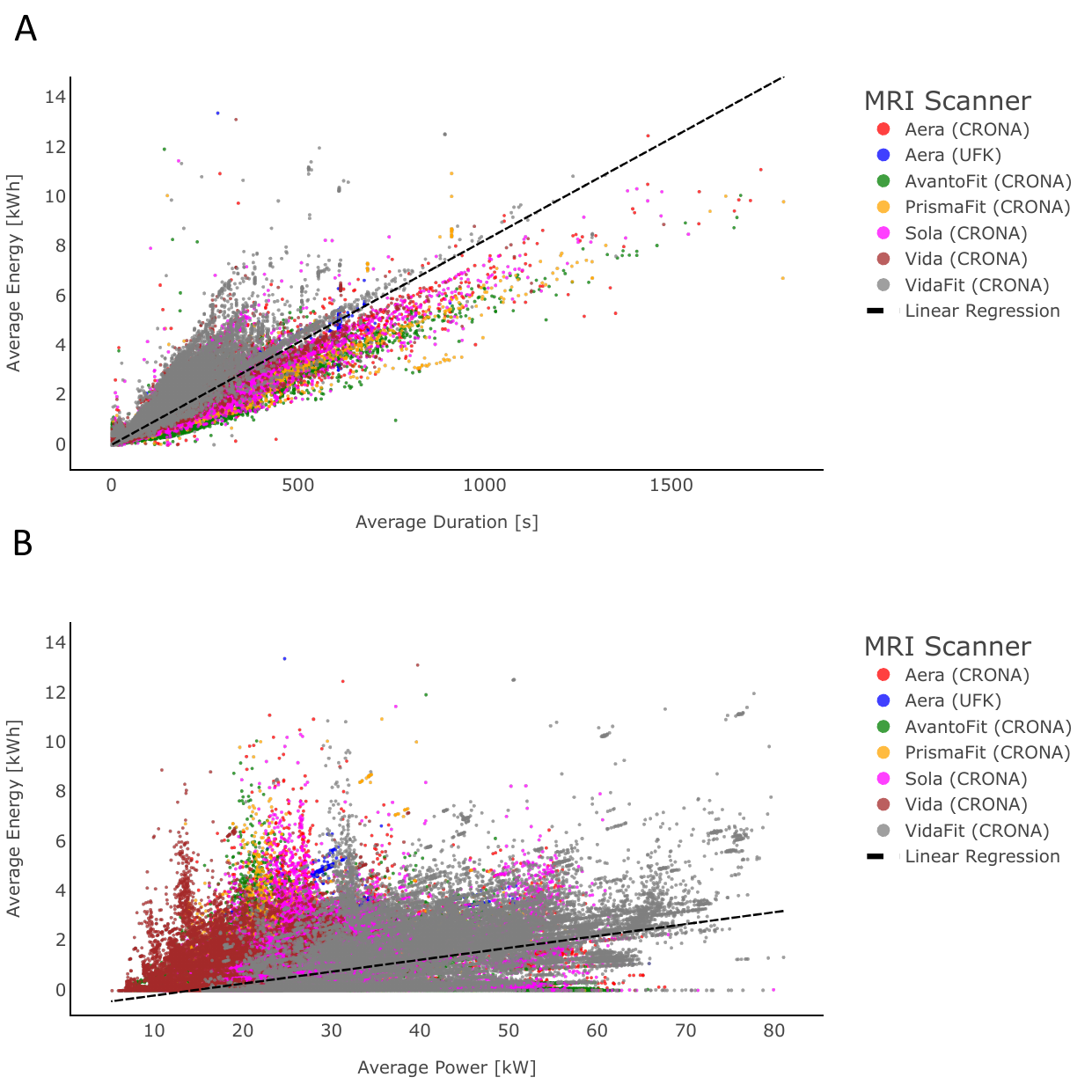}
    \caption{Scanner-specific scatter plots were used to illustrate the relationship between the energy consumption of individual MRI sequences and their average scan duration and active power consumption. A: Relationship between average scan duration and average energy consumption. Linear regression demonstrated a strong positive relationship between scan duration and energy consumption ($y = 0.008x + 0.020$, $R^2 = 0.769$) over all scanners. For improved visualization, the y-axis limits were determined based on the regression results, resulting in the exclusion of outliers. B: Relationship between average active power and average energy consumption. A weaker positive relationship was observed based on linear regression ($y = 0.048x - 0.669$, $R^2 = 0.426$). The same y-axis limits as in A were used to ensure comparability between the two plots.}
    \label{fig:figure7_relation_energy_scan_duration_power}
\end{figure}

Analysis of the relationship between average energy consumption and average scan duration for individual MRI sequences revealed a strong linear relationship over all MRI scanners ($y = 0.008x + 0.020$, $R^2 = 0.769$). Several groups of sequences followed approximately linear patterns, indicating similar relationships between scan duration and energy consumption within these groups. In contrast, vertical clusters represented sequences with similar scan durations but varying energy consumption, suggesting that differences in scanner power demand and acquisition parameters also contribute to energy variability.  (Fig. \ref{fig:figure7_relation_energy_scan_duration_power} A). \\

Conversely, the relationship between average power consumption and energy consumption was considerably weaker. A linear regression model showed a moderate positive relationship ($y = 0.048x - 0.669$, $R^2 = 0.426$). Although energy consumption generally increased with average power, the data showed a relatively homogeneous distribution across power levels, with substantial variability in energy consumption at comparable power values. A distinct near-zero band of energy consumption was observed, with numerous measurements exhibiting very low energy consumption despite a wide range of average power values. Whereas at similar power levels, energy consumption showed substantial variability (Fig. \ref{fig:figure7_relation_energy_scan_duration_power} B). \\

\begin{figure}[t]
    \centering
    \includegraphics[
        width=\linewidth,
        keepaspectratio
    ]{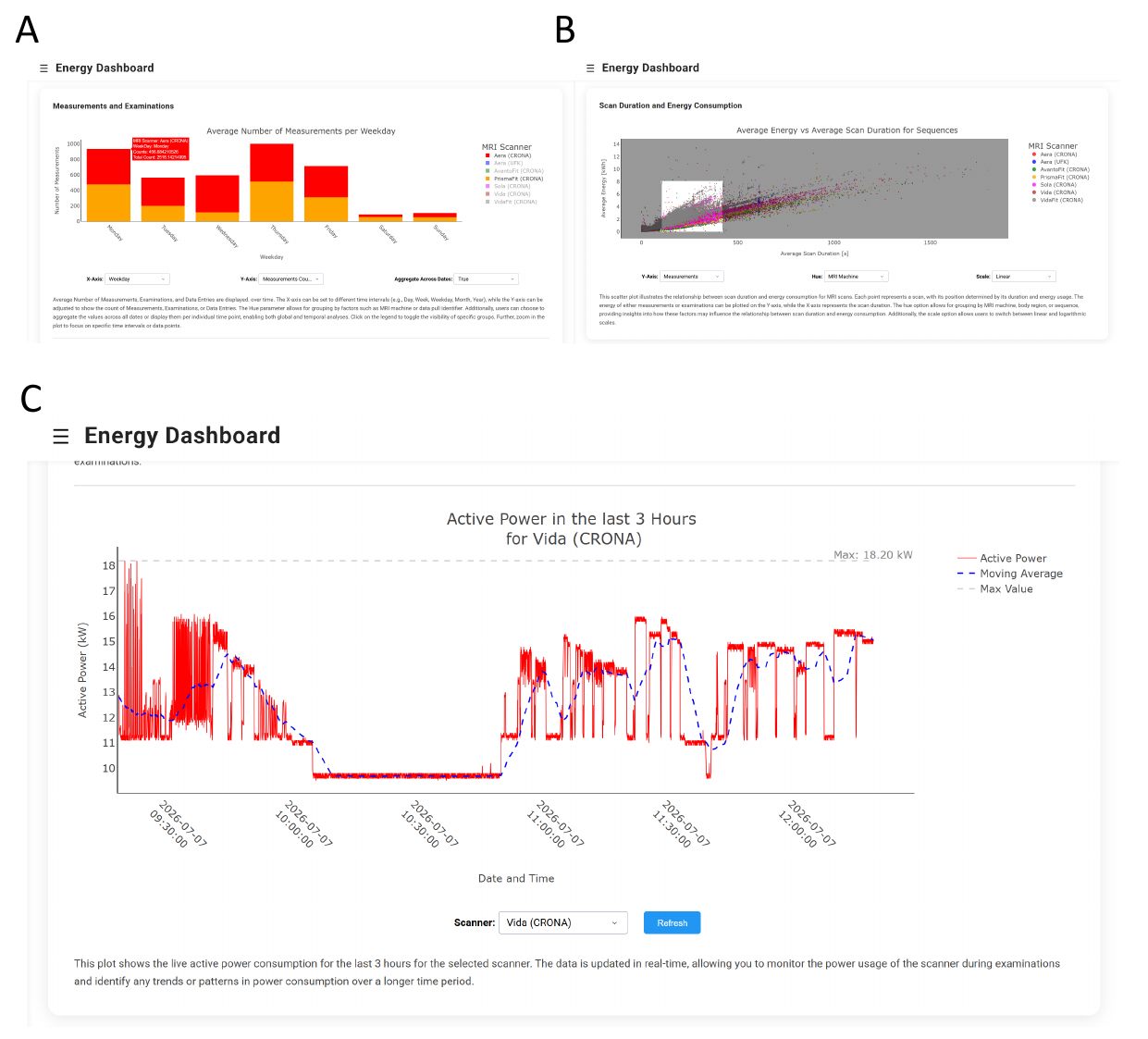}
    \caption{Screenshots of the hosted energy dashboard. (A) Energy consumption for each measurement. The buttons below the plot allow users to individually adjust the aggregation of the displayed data. Hovering over the measurements provides additional information, while clicking on entries in the legend allows specific groups to be selected and analyzed. (B) Relationship between energy consumption and scan duration for the individual sequences. By selecting a specific area within the plot, users can zoom in on that region and investigate the corresponding sequences in greater detail. (C) Live data streamed directly from the hospital, showing the power consumption of various MRI scanners over the last three hours.}
    \label{fig:figure8_energy_dashboard}
\end{figure}

\section{Discussion}
\label{discussion}
This study extends previous analyses of MRI energy consumption by linking energy recordings with scanner utilization, examination characteristics, scanning protocols, and acquisition parameters. In contrast to previous studies \cite{Heye_2020, Heye_2023, Woolen_2023, Afat_2025}, which have largely focused on individual aspects of scanner operation or specific energy-saving interventions, our approach systematically connects energy demand across different levels of analysis across seven MRI scanners over 1.5 years. This integration enables a more granular characterization of MRI energy demand and provides a comprehensive baseline for identifying the factors that drive energy consumption. The resulting dataset and analysis provide a foundation for future data-driven approaches to energy optimization and the development of effective energy-saving strategies. \\

The high number of daily measurements resulted from the inclusion of adjustment and localizer scans, which contribute to overall MRI energy consumption through tasks such as field adjustment, RF calibration, coil sensitivity estimation, and examination planning. Furthermore, missing data occurred due to the reliance on a continuous and stable network connection for recording temporal energy data throughout the day. Network interruptions during data collection, therefore, resulted in incomplete measurements and subsequent data loss. \\

Clear differences in average energy consumption were observed between scanners, examinations, and individual measurements. In particular, the relative energy consumption among scanners varied depending on whether it was analyzed at the examination or measurement level. This suggests that examination-level energy consumption is determined not only by the energy demand of individual measurements but also by the number and composition of measurements included in each examination protocol. Temporal fluctuations in energy consumption were also observed and are likely related to structured clinical scheduling patterns, where specific examination types, such as neurological or cardiac imaging programs, are assigned to predefined time slots or days of the week. Consequently, daily energy consumption varies depending on the scheduled examinations, with lower demand during weekends due to fewer routine scans. Last, the elevated energy consumption of the Vida scanner (CRONA) on Saturdays is likely explained by overnight research scans. \\

Variations in energy consumption are closely related to differences in scanner occupancy. Scanner utilization varied substantially between MRI systems, reflecting differences in patient load, operating hours, clinical workflows, scanner specialization, and available licenses. In particular, the PrismaFit (CRONA) scanner showed substantial fluctuations in occupancy. This pronounced variation may be attributed to its use as a dedicated research scanner, resulting in more variable utilization patterns than scanners used primarily for clinical examinations. Importantly, occupancy can be considered an indicator of patient load and may provide insight into how the clinical workload is distributed across available scanners. Similar temporal patterns to those observed for energy consumption were present, likely due to the same underlying causes. \\

Analyzing energy consumption across operating modes provided insight into the distribution of energy demand between active scanning, idle operation, and eco-power mode, and its temporal variations. Such information is valuable for estimating both baseline and peak power requirements, which may support infrastructure planning, such as the design of electrical supply systems. As expected, idle and eco-power mode consumption remained relatively constant throughout the week due to continuous cooling requirements. Notably, these modes accounted for more cumulative energy consumption than active scanning, suggesting that reducing energy consumption during non-scanning periods may offer great optimization potential without affecting image acquisition. When comparing scanners, energy consumption across operating modes should be interpreted in relation to scanner occupancy, as higher utilization leads to increased energy demand from active scanning. Nevertheless, occupancy alone does not fully explain the observed differences between scanners, indicating that additional factors contribute to energy consumption. Future work should therefore consider developing normalized efficiency metrics that incorporate scanner utilization to enable more meaningful comparisons between MRI systems. \\

Grouping energy consumption by body region provided a broad overview but was insufficient to explain variations in energy demand, as examinations within the same anatomical region can involve different sequences and acquisition parameters. Some of the observed differences may be partly attributed to variations in scan durations. Future analyses should therefore focus on sequence-based grouping using data-driven approaches to identify sequences with similar energy consumption patterns and optimization potential. However, this remains challenging because sequence names alone are often inconsistent and do not fully capture acquisition characteristics, requiring approaches that incorporate both sequence information and acquisition parameters. \\

Based on the physical relationship between energy, power, and time ($E = P \cdot t$), scan duration and average power were expected to influence the energy consumption. This was partially supported by the strong linear relationship observed between scan duration and energy consumption ($R^2 = 0.769$). In comparison, average power consumption showed a weaker relationship with energy consumption ($R^2 = 0.426$), indicating a smaller contribution to the observed variation in energy demand. \\

The approximately linear patterns in the energy–scan duration plot represent groups of sequences in which longer scan durations were associated with higher energy consumption while power demand remained relatively similar. In contrast, vertical clusters represent sequences with similar scan durations but different energy consumption, suggesting that differences in power demand also contribute to energy variability. \\

The energy–power relationship also showed a distinct near-zero band, with very low energy consumption observed across a wide range of average power values. This pattern may be explained by sequences with very short scan durations, for which the limited acquisition time resulted in low overall energy consumption despite differences in average power demand. Thus, variations in power demand had a comparatively small effect on total energy consumption for these short acquisitions. Simultaneously, substantial variability in energy consumption was observed even at similar power levels, indicating that average power alone cannot adequately explain the observed differences in energy demand. This further highlights the importance of scan duration and other sequence-specific factors in determining overall energy consumption. \\

Overall, the results indicate that scan duration is an important factor associated with energy consumption, while differences in power demand may contribute to additional variability. Future analyses should therefore focus on identifying the underlying acquisition parameters that determine scan duration and scanner power demand, as these measures represent higher-level features that may reflect the combined effects of more granular parameters. \\

Lastly, the complexity of the dataset prompted the development of an interactive dashboard to support the exploration and visualization of energy consumption patterns and relationships across the different analyses. The dashboard enables interactive aggregation and filtering of observations, exploration of individual analyses, and zooming into specific regions of the data. In addition, it incorporates a live data connection from the hospital, enabling continuous visualization and real-time monitoring of MRI scanner power consumption over the preceding three hours (Fig. \ref{fig:figure8_energy_dashboard}). This interactive framework facilitates the identification of relevant patterns and provides a basis for further investigation of MRI energy consumption behavior. \\

We acknowledge several limitations of this study. First, the analysis was conducted retrospectively in a clinical setting. While this approach was intentionally chosen to characterize scanner energy consumption under real-world conditions, it inherently introduces variability in scanner utilization, examinations, and patient populations. Consequently, the findings provide broad insights into energy consumption patterns, while more detailed conclusions that would require standardized experimental conditions were avoided. Second, all seven MRI scanners were from the same manufacturer and energy measurements were conducted at a single hospital, which may limit the generalizability of the findings to other scanner manufacturers and clinical settings. Third, energy consumption depends on the specific examinations performed, as differences in examination protocols, sequence composition, and scan duration can substantially affect the resulting energy demand. For example, \citet{Heye_2020} reported an average energy consumption of 19.9 kWh for an examination, whereas our analysis yielded 13.50 kWh. This difference highlights the need for standardized benchmarks for comparing MRI energy consumption across scanners, institutions, and studies. Finally, the analysis primarily relied on linear methods to investigate relationships between energy consumption and relevant features. While these methods provide interpretable measures of association, they may not capture nonlinear or more complex dependencies between acquisition parameters and energy demand. \\
      
Future work should include recordings under more controlled conditions to enable a more detailed investigation of the factors influencing MRI energy consumption. High-resolution live data acquisition could provide a better understanding of temporal energy patterns and enable previous findings to be validated under standardized conditions. A better understanding of these factors would also enable the development of standardized benchmarks for assessing energy efficiency and identifying opportunities to optimize scanner operation. Finally, future studies should apply more advanced modeling approaches that can account for nonlinear relationships and interactions between acquisition parameters, scan duration, power demand, and energy consumption. \\

\section{Conclusion}
\label{conclusion}
In conclusion, merging detailed MRI sequence and acquisition parameters with their corresponding energy recordings enables a more comprehensive characterization of MRI energy consumption beyond the examination and scanner levels. By combining different analytical steps, the study provides an integrated view of energy demand rather than considering individual aspects of scanner operation in isolation, an approach that was particularly enabled by the energy dashboard. The results demonstrate that scan duration is an important factor related to energy consumption, while variations in power demand contribute to a lesser extent. Further investigations using controlled recordings, high-resolution temporal data, and data-driven modeling approaches are needed to identify the specific acquisition parameters that drive energy consumption beyond high-level features such as the aforementioned scan duration and average power demand. These insights can support the development of meaningful energy-efficiency metrics and standardized benchmarks, as well as targeted strategies for optimizing scanner operation and software.
\bibliographystyle{unsrtnat}
\bibliography{references}  






\end{document}